\documentclass{article}

\usepackage{amssymb,amsmath}
\usepackage{fullpage}
\usepackage{graphicx}
\usepackage{amsmath}		
\usepackage[margin=1.0in]{geometry}
\usepackage{setspace}
\usepackage{color}
\usepackage{fancyhdr}
\usepackage{collcell}
\usepackage{datatool}
\usepackage{environ}
\usepackage{latexsym}
\usepackage{amssymb}
\usepackage{epsfig,amsmath,graphics}
\usepackage{epstopdf}
\usepackage{verbatim}
\usepackage{wasysym}
\usepackage{feynmp-auto}
\usepackage{authblk}
\usepackage{xcolor}
\usepackage{enumitem}
\usepackage[utf8]{inputenc}
\usepackage{slashed}
\usepackage{cite}

\usepackage{empheq}

    \newlength\fsep
    \newsavebox\widebox

\usepackage[skip=0pt]{caption}

\title{Dark Energy Radiation}
\date{\today}

\author[1,2]{Kim V.~Berghaus}	
\author[3]{Peter W.~Graham}
\author[1]{David E.~Kaplan}
\author[4]{Guy D.~Moore}
\author[1]{Surjeet Rajendran}
\affil[1]{Department of Physics \& Astronomy, The Johns Hopkins University, Baltimore, MD  21218, USA}
\affil[2]{C.N. Yang Institute for Theoretical Physics, Stony Brook University, NY 11794, USA}
\affil[3]{Stanford Institute for Theoretical Physics, Department of Physics, Stanford University, Stanford, CA 94305, USA}
\affil[4]{Institut f\"{u}r Kernphysik, Technische Universit\"{a}t Darmstadt,
Schlossgartenstra{\ss}e 2, D-64289 Darmstadt, Germany}

\begin{document}

\maketitle

\begin{abstract}
We show that if dark energy evolves in time, its dynamical component could be dominated by a bath of dark radiation.  Within current constraints this radiation could have up to $\sim 10^4$ times more energy density than the cosmic microwave background.
We demonstrate particular models in which a rolling scalar field generates different forms of dark radiation such as hidden photons, milli-charged particles and even Standard Model neutrinos.
We find the leading effect on the late-time cosmological expansion history depends on a single parameter beyond $\Lambda$CDM, namely the temperature of the dark radiation today.
Cosmological observations of this modified expansion rate could provide a striking signature of this scenario.
The dark radiation itself could even be directly detectable in laboratory experiments, suggesting a broader experimental program into the nature of dark energy.
\end{abstract}

\tableofcontents

\section{Introduction}
\label{Sec:Introduction}
A variety of cosmological measurements have established that the universe is currently undergoing accelerated expansion driven by dark energy. This dark energy dominates the energy density of the universe. It does not clump into galaxies and its evolution is constrained to be slower than that of the other components of the universe.  Notably, these properties have been established by cosmological measurements that are sensitive only to the gravitational effects of dark energy. Much like the case of dark matter, it is likely that the properties of dark energy can be more deeply probed if the dark energy has non-gravitational interactions with the standard model. How can the dark energy couple to the standard model and what kinds of signals can these interactions lead to? A systematic identification of these could lead to a robust program of laboratory based probes of dark energy, similar to the effort that is currently underway to identify the nature of dark matter. 

Unlike dark matter, such a program has not been widely pursued for dark energy due to the widespread theoretical assumption that the dark energy is a cosmological constant. Indeed, a cosmological constant is the simplest proposal for dark energy and such a constant cannot be probed in laboratory experiments. But this is simply theory prejudice - the experimental nature of physics requires observational probes to establish if the dark energy is actually a constant or if it has a dynamical component, {\it e.g.}, a kinetic energy. This kinetic energy can couple to the standard model through non-gravitational interactions and could potentially be probed in the laboratory. In fact, there are significant theoretical motivations to search for such a kinetic energy.  The cosmological constant is observationally known to be orders of magnitude smaller than identifiable standard model contributions to it and thus appears to require a massive fine tuning. At present, the only known mechanisms ({\it e.g.}, \cite{Abbott:1984qf,  Banks:1984tw, Steinhardt:2001st, Alberte:2016izw, Graham:2019bfu, Graham:2017hfr})
to naturally explain this fine tuning requires dynamical evolution, implying the existence of a dynamical component of dark energy. 

Given the isotropic nature of dark energy, the simplest model of dynamical dark energy is a light scalar field that is slowly rolling down its potential. While the mass of this field has to be smaller than the Hubble scale ($\sim 10^{-43}$ GeV) today, its kinetic energy could be significantly larger --- constrained to be no larger than a few percent of the dark energy density today $\sim \text{meV}^4$. The coupling of this kinetic energy to the polarization of light (see e.g.~\cite{Boyle:2001du, Carroll:1989vb, Minami:2020odp} and references)
and the spin of nucleons/electrons has been investigated (see e.g.~\cite{Pospelov:2004fj, Romalis:2013kkt, Brown:2010dt, Graham:2020kai}). In these, the dark energy provides a homogeneous background and the motion of the test particle against the background gives rise to polarization rotation/spin precession. The spatial/temporal coherence and the frequency of these signals are all set by the Hubble scale today {\it i.e.} the age of the universe. Experimentally, while the extended temporal coherence is useful to combat sensor noise, the DC nature of the signal gives rise to challenging systematics. 

In this paper, we point out that the phenomenology of dynamical dark energy is richer than simply the motion of test particles against a homogeneous background, giving rise to qualitatively different experimental signatures. Specifically, if the scalar field has  technically natural interactions with a dark Yang-Mills sector, these interactions can give rise to additional friction/dissipation channels for the kinetic energy of the scalar field. These dissipation mechanisms can convert the coherent and homogenous kinetic energy of the field into hot radiation. The energy density in this radiation can be as large as a few percent of $\text{meV}^4$.  While this is considerably larger than the energy density set by $n_{\text{eff}}$ measurements from the CMB (in fact, this energy density is larger than that of the CMB today), there is no conflict with the CMB since this radiation occurs from the conversion of dark energy today rather than in the early universe. 

If the dark energy were to have this phenomenology, there are considerable experimental implications. First, these dissipation mechanisms can remove a significant chunk of the kinetic energy of the field leading to null results in some of the DC experimental probes discussed above. 
In particular the spin precession measurements in laboratories today would have a greatly reduced signal.  However, searches for polarization rotation of the CMB (cosmic birefringence) would still be important as even a tiny integrated kinetic energy would be detectable beyond astrophysical limits.
Second, the conversion of the kinetic energy into hot radiation could be detected indirectly through cosmological measurements of the expansion rate.  Third, this conversion could be detected directly by searching for this radiation to interact with a detector. This requires new kinds of experiments. The hot radiation is at a high frequency $\sim$ THz ($\sim$ meV) where systematic backgrounds are under better control than a DC experiment. But, the signal is not coherent and thus requires enhanced sensing. 

The properties of the standard model imply that this radiation is not in the form of photons. The high mass of the electron relative to the dark energy scale as well as astrophysical limits  on the direct coupling of light scalar fields to electromagnetism makes the conversion into photons inefficient. In this paper, we point out that this conversion can be efficient through Yang-Mills sectors, hidden/dark photons, milli-charged particles and  right handed neutrinos. 

The rest of this paper is organized as follows. In section \ref{sec:scalarfriction}, we outline the basic technique that allows the kinetic energy of dark energy to be converted into radiation in a technically natural way. We then show how this radiation can manifest itself in the form of pure Yang-Mills gauge bosons (section \ref{sec:yangmills}), hidden photons (section \ref{sec:hiddenphotons}), milli-charged particles (section \ref{sec:hiddenphotons}) and right handed neutrinos (section \ref{sec:neutrinos}). Following this, in section \ref{sec:cosmology} we discuss cosmological probes of this scenario and conclude in section \ref{sec:discussion} where we comment on various direct detection possibilities. 

\section{Friction and Radiation}
\label{sec:scalarfriction}

The point of this section is to show that dark energy could easily have a radiation component consisting of a variety of possible particle types. 
We focus on models in which a rolling scalar field couples to such a light sector.  The rolling scalar populates the light sector, which in turn thermalizes and creates friction for the rolling scalar as a form of back-reaction.

We take arguably the simplest model of a rolling scalar field, one with a linear potential $V(\varphi) = - C\varphi$.  The equation of motion for such a field will be
\begin{equation}
    \label{generalEOM}
    \ddot{\varphi} + 3 H \dot{\varphi} + \Upsilon\dot{\varphi} = C
\end{equation}
Where $H$ is the Hubble rate and $\Upsilon$ is the (generally temperature-dependent) source of friction that we will exploit in this paper.  The over-dot denotes time derivatives in standard FRW coordinates.  We have chosen the origin of $\varphi$ to be the point of zero vacuum energy.  Observation requires $\varphi<0$ today (positive vacuum energy) and we will look for solutions where $\ddot{\varphi} < (3H +\Upsilon) \dot{\varphi}$.  Unless otherwise noted, we will assume $\Upsilon \gg 3 H$.  Thus dark energy density will consist of vacuum energy plus a quasi-thermal bath and a near-negligible contribution from $\varphi$'s kinetic energy.

In the rest of this section, we present minimal models that can generate different particle contents in the background dark radiation.

\subsection{Non-Abelian Gauge Sector}
\label{sec:yangmills}

For a non-Abelian gauge theory with fermions and an axion-like coupling of $\varphi$ to the gauge fields, we have the Lagrangian
\begin{equation}
  \label{genericL}
  -{\cal L} = \frac{1}{2g^2} \:\mathrm{Tr}\: G_{\mu\nu} G^{\mu\nu}
  + \bar\psi \left( \slashed{D} + m \right) \psi
  + \frac 12 \partial_\mu \varphi \partial^\mu \varphi
  + \frac{\varphi}{f} \frac{\:\mathrm{Tr}\:
    G_{\mu\nu} \tilde{G}^{\mu\nu}}{16\pi^2}
  - C\varphi
\end{equation}
where we take a gauge group $SU(N_c)$ and the $N_f$ fermions $\psi$ are in some representation $R$ of the gauge group.  Contributions to dynamical friction on a pseudo-scalar from a QCD-like sector has been of significant interest as it affects  the calculation of the axion dark-matter abundance \cite{McLerran:1990de}.

First we investigate the simplest case, pure Yang-Mills (essentially the $m\rightarrow\infty$ limit).  The field equation of motion is
\begin{equation}
  \label{scalarvev}
  \partial_\mu \partial^\mu \varphi = -C
  + \frac{\mathrm{Tr}\: G_{\mu\nu} \tilde{G}^{\mu\nu}}{16\pi^2 f} .
\end{equation}
The last term is related to the Chern-Simons number density.  When a thermal average is taken of this equation, the last term will contribute to rolling friction due to real-time sphaleron processes when $\dot\varphi\neq 0$.  This contribution is
\begin{equation}
  \label{ffdualYM}
  \left\langle
  \frac{\mathrm{Tr}\: G_{\mu\nu} \tilde{G}^{\mu\nu}}{16\pi^2}
  \right\rangle = \frac{\Gamma_{\mathrm{sph}}}{2T}
  \left( \frac{\dot\varphi}{f} \right)
\end{equation}
where the sphaleron rate is of order
\begin{equation}
  \label{sphaleron-est}
  \Gamma_{\mathrm{sph}} \sim N_c^5 \alpha^5 T^4
\end{equation}
up to logs of $\alpha$.  The factor $1/2T$ is the relation between the free diffusion rate for topological charge $\Gamma_{\mathrm{sph}}$ and the
rate at which a chemical potential drives topological transitions,
(see Ref. \cite{Moore:1996qs}).  For a discussion of the sphaleron rate see
\cite{Arnold:1996dy,Bodeker:1998hm,Moore:2010jd}, and for a general discussion of
chemical potentials, particle number violation, and sphaleron
processes, see
\cite{Kuzmin:1985mm,Giudice:1993bb,Rubakov:1996vz}.
Thus the friction coefficient in \eqref{generalEOM} is \cite{Berghaus:2019whh} $\Upsilon \sim (N_c \alpha)^5 T^3 /f^2$.  This friction beats Hubble friction when $f\ll T \sqrt{T/H} (N_c \alpha)^{5/2}$.

Now we can estimate the temperature as a function of the remaining parameters.  The dark radiation density, $\rho_{DR}$, will lose energy density via  redshift from the expanding universe,  and gain energy density from the friction's back reaction:
\begin{equation}
\label{DRevolve}
 \dot{\rho}_{DR}  = - 4H \rho_{DR} +  \Upsilon \dot{\varphi}^2 
 \end{equation}
 where $\rho_{DR} = (2N_c^2-1)(\pi^2/30)T^4$, accounting for
 the gluons and the thermalized axions (thermal excitations of the inflaton field) which accompany them.
 We can plug in for $\dot{\varphi}$ by solving for it in \eqref{generalEOM} in the slow-roll ($\ddot{\varphi}\rightarrow 0$) and thermal friction ($\Upsilon\gg H$) limits, giving $\dot{\varphi}\simeq C/\Upsilon$.  If we also assume quasi-steady-state behavior in \eqref{DRevolve}, $\dot{\rho}_{DR}\sim 0$, we can solve for $T$:
 \begin{equation}
 \label{Tsteady}
   T \sim  \left(\frac{C^2 f^2}{H N_c^7 \alpha^5}\right)^{1/7} .
\end{equation}
    The temperature's slow ($-1/7$ power) dependence on $H$ shows this is a reasonable approximation.  However, in Section \ref{sec:cosmology}, we solve the exact equations of motion analytically and numerically.

Note, we can choose parameters such that $T\sim $ meV. For thermal friction dominance, this amounts to $f\ll$ TeV.  As this is all occurring in a hidden sector it is not subject to collider bounds.  When one includes couplings to standard model particles, there exists  a broad range of parameter space allowed by current experiments.

Also note,  the following assumptions are required to make our analysis  valid.
We assume that $\dot\phi$ is slowly varying on the $1/\alpha^2 T$ time scale on which the thermal medium reacts to these effects.  If this is not the case, then the response is more complicated and is not determined by the sphaleron rate. However, this is a trivial constraint.
We also assume that the $\dot\varphi G \tilde{G}/f$ term is smaller than the other terms in the YM equations of motion so we can linearize in it in the usual fluctuation-dissipation sense.  This happens as long as  $\dot\varphi/f<T$, which is again trivial to satisfy.  Finally, we take  $\alpha^2 T > H$ so the Hubble rate is small compared to the rate of thermalization, which simply requires $\alpha$ to not be too small.

 Lastly, note that this quasi-steady-state behavior is relatively insensitive to initial conditions. Setting $\dot{\varphi} = 0$ and $T=0$ at early times still asymptotically generates this friction and heat bath today.  A large $T$ at early times will red-shift away and is anyway limited by CMB sensitivity to additional degrees of freedom.  A large $\dot{\varphi}$ at early times redshifts even faster than radiation and is then damped by thermal dynamics.  Thus, the solution we find is an attractor.

\subsubsection{Non-Abelian with Fermions}

The existence of fermions in this sector can suppress the friction as sphaleron transitions can build up chiral charge and associated chemical potential, which suppresses the sphaleron rate.  A fermion mass violates the chiral symmetry and thus reduces the chiral charge.  These issues only occur if the fermion masses are very light; 
we will see below that, if the fermion mass $m$ grows larger than $\alpha^2 T$, the sphaleron rate approaches that of the pure YM case.

If the mass is zero, there are two conserved fermionic numbers,
the left-handed fundamental number $n_F$ (for fermions in the fundamental representation) and the left-handed
anti-fundamental number $n_{\bar F}$, with associated currents
$j^\mu_F$ and $j^\mu_{\bar F}$.
If we think of the two Weyl fermions as one Dirac fermion,
$n_F$ is the net left-particle number and $n_{\bar F}$ is minus the
net right-particle number.  
We will assume initially
$n_F=n_{\bar F}=0$.
Again in the absence of a mass, each current obeys
\begin{equation}
  \label{nfviolation}
  \partial_\mu j^\mu_F =
  \frac{\mathrm{Tr}\: G_{\mu\nu} \tilde{G}^{\mu\nu}}{16\pi^2}
  =  \partial_\mu j^\mu_{\bar F}
\end{equation}
where the field strength trace is the topological charge density.  For
a general representation $R$ we would find
\begin{equation}
  \label{genrep}
  \partial_\mu j^\mu_R = 2 T_R
  \frac{\mathrm{Tr}\: G_{\mu\nu} \tilde{G}^{\mu\nu}}{16\pi^2}
\end{equation}
where $T_R = C_R d_R / d_A$ is the trace normalization of the
representation ($1/2$ for fundamental, $N_c$ for adjoint, etc),
$C_R$ is the Casimir, $d_R$ is the dimension, and $d_A=N_c^2-1$ is the
dimension of the adjoint representation.  Assuming a statistically uniform system, this reduces to equations for particle number, such as
\begin{equation}
  \label{nfevolve}
  \partial_t n_R = 2T_R
    \frac{\mathrm{Tr}\: G_{\mu\nu} \tilde{G}^{\mu\nu}}{16\pi^2} .
\end{equation}

If the particle number is otherwise conserved (or for small mass, long
lived), then the fermions otherwise equilibrate, and the particle
number is related to a chemical potential $\mu_R$ via
\begin{align}
  \label{mu-and-n}
  n_R(\mu) & = d_R \int \frac{d^3 p}{(2\pi)^3} \left(
  \frac{1}{e^{(p-\mu_R)/T}+1} - \frac{1}{e^{(p+\mu_R)/T}+1} \right)
  \simeq \frac{d_R \mu_R T^2}{12} \,,
  \\ \nonumber
  \mu_R(n) & = \frac{12n_R}{d_R T^2} \,.
\end{align}
Here the two terms in the first line are the particles and the
antiparticles, and we have expanded to linear order in $\mu_F$.

Now consider the equations of motion for the Chern-Simons number itself.  Besides the driving force $\dot\varphi$ acting on it, there is also
the energy cost of producing particle number against a chemical
potential:
\begin{equation}
  \label{ffdual}
  \left\langle
  \frac{\mathrm{Tr}\: G_{\mu\nu} \tilde{G}^{\mu\nu}}{16\pi^2}
  \right\rangle 
  = \frac{\Gamma_{\mathrm{sph}}}{2T}
  \left( \frac{\dot\varphi}{f} - \sum_R 2T_R \mu_R \right)
\end{equation}
where we showed the general situation for multiple fields
in different representations $R$.  The factor $2T_R$ is the number of
particles produced by a sphaleron event, each of which costs a free
energy $\mu_R$.

Taking the thermal average of equations of motion \eqref{generalEOM} and \eqref{nfevolve}, and inserting the relations
for the chemical potential \eqref{mu-and-n}, we find the following combined equations (for a massless fermion in representation $R$):
\begin{eqnarray}
  \label{phi-evolve}
  - \ddot{ \varphi }
   = -C
  + \frac{\Gamma_{\mathrm{sph}}}{2Tf} \left( \frac{\dot \varphi}{f}
  -  \frac{24 T_R n_R}{d_R T^2} \right) ,
  \\
  \label{nF-evolve}
  \partial_t n_R = \frac{T_R \Gamma_{\mathrm{sph}}}{T} \left(
  \frac{\dot \varphi}{f} -  \frac{24 T_{R} n_{R}}{d_{R} T^2} \right) .
\end{eqnarray}
We see that there is no solution where $\ddot\varphi$ vanishes asymptotically; instead, the particle number grows and $\varphi$ continues to accelerate.  As a side note, there is still effective dissipation even in this exactly massless case.  These equations \eqref{phi-evolve} and \eqref{nF-evolve} can be solved for an asymptotic solution which still has energy being dumped from the scalar field into the thermal bath.  This rate of energy transfer is highly suppressed and we will not focus on these solutions.

However, in the case of a small non-zero fermion mass ($m\neq 0$), there are chirality-violating scattering
processes, with a matrix element proportional to the mass at leading order.  This means
that there is an additional term in the particle number evolution
equation:
\begin{equation}
  \label{nF-evolve2}
  \partial_t n_R = \frac{T_R \Gamma_{\mathrm{sph}}}{T} \left(
  \frac{\dot \varphi}{f} -  \frac{24 T_{R} n_{R}}{d_{R} T^2} \right)  - \kappa \frac{n_R N_c \alpha m^2}{T^2}
\end{equation}
with $\kappa$ a coefficient which one could calculate in principle and should be ${\cal O}(1)$ \cite{Boyarsky:2020cyk}.
Now we see that $n_R$ can approach a constant value.  Setting its time dependence in \eqref{nF-evolve2} to zero and solving for $n_R$, one can now generate an effective friction rate $\Gamma_\text{eff}$ in the $\varphi$ equation of motion
\begin{equation}
   - \ddot{\varphi} =  - C + \Gamma_\text{eff}\, \dot{\varphi} 
\end{equation}
with
\begin{equation}
    \Gamma_\text{eff} = \frac{\Gamma_{\mathrm{sph}}}{2 T f^2} \left(
        \frac{\Gamma_{ch}}{\Gamma_{ch} + \frac{24 T_R^2}{d_R T^3}\Gamma_{\mathrm{sph}}}\right)
\end{equation}
where we defined the chirality-violation rate $\Gamma_{ch} \equiv \kappa N_c \alpha m^2/T$.   If the $\Gamma_{ch}$  term is larger than the sphaleron term --  {\it i.e.}, if $m^2/T^2 \gg (24 T_R^2/d_R\kappa)(N_c \alpha)^4$ -- then the particle number relaxes primarily due to spin flips before it can back-react on the sphaleron processes, and thus for a large enough mass, the original friction is restored. 
 For a finite mass, friction at some level is maintained and steady-state solutions are possible. In this limit where $\Gamma_{ch}$ is small, the thermal friction beats Hubble friction if $\Gamma_\text{eff}\simeq (\kappa d_R N_c \alpha)/(48 T_R^2) (T m^2/f^2) > H$, and thus it is possible to have significant dark energy radiation for a wide range of masses.

\subsection{Dark Photons and Milli-Charged Particles}
\label{sec:hiddenphotons}

There are a number of ways to make $U(1)$ gauge bosons (dark photons) part of the dark energy thermal bath.  The simplest way is to take the QCD-like model above and charge the fermions under an unbroken $U(1)$.  Take the new fine-structure constant to be $\alpha'$.  In the thermal bath, the rate of dark photon production should be at least $\Gamma_{\gamma'} \sim \alpha \alpha' T$.  As long as this is much larger than the Hubble expansion rate, a thermal abundance of dark photons would be generated.

One does not need fermions in the non-Abelian sector to thermalize dark photons.  Additional ways to thermalize light $U(1)$s is through higher-dimensional operators {\it e.g.},
\begin{equation}
\label{phi-darkphoton}
        {\cal L}\supset \frac{\varphi}{f'}F'^{\mu\nu}{\tilde F'}_{\mu\nu}
\end{equation}
where $F'$ is the field strength of a dark photon.
Because the YM to dark photon cross section now goes as $T^2/(f'f)^2$, a much lower $f', f$ would be required to thermalize this sector, roughly $f' \sim f \lesssim 10$ keV for $T\sim$ meV.  The only other requirement is that the mass of the dark photon, $m_{A'}$ satisfies $T > m_{A'} \gg H$ (we discuss the massless case below).

This thermalization process can be sped up if there are interactions in the $U(1)$ sector.  This is because when thermal friction dominates, the rolling dark energy field is all but guaranteed to thermalize as well.  Roughly, if the rate $\Gamma_{GG\rightarrow G\varphi}\simeq \sigma n_G \sim (\alpha T^3/f^2) > H$, then $\varphi$ particles are also thermal.
Once a small amount of dark photons is produced via \eqref{phi-darkphoton}, a runaway process can produce a thermal abundance in this sector.  For example, a sector of dark photons $A'_\mu$ and dark-charged fermions $\chi$ will have  scattering cross sections $\sigma_{\varphi\chi\rightarrow A'\chi}$ and $\sigma_{\varphi A'\rightarrow \chi\chi}$ of order $\alpha'/f'^2$.  The growth rate of the number density of hidden sector particles $X\equiv A',\chi$ would be
\begin{equation}
    \frac{dn_X}{dt}\sim n_X n_\varphi \sigma_{\varphi X\rightarrow XX} \sim n_X \frac{\alpha' T^3}{f'^2}.
\end{equation}
If the characteristic rate $\frac{\alpha' T^3}{f'^2}\gg H$ (up to a log factor), a thermal abundance of dark photons and dark-charged fermions should be produced.  If the dark photons kinetic-mix with regular photons, this would also constitute an abundance of milli-charged fermions.

For completeness, we note that in the case of an extremely light or massless dark photon without charged fermions, an abundance of very long wavelength modes is also generated by a tachyonic instability, and in fact becomes a new source of friction for $\varphi$ \cite{Campbell:1992hc}. Via the coupling \eqref{phi-darkphoton}, a population of vectors will be produced when $\dot{\phi}/f > m_{A'}$ due to an instability in the mode equation for the vectors  \cite{Campbell:1992hc, Anber:2009ua, Hook:2016mqo}:
\begin{equation}
\label{eq-sorbo}
\ddot{A'_{\pm}} + \left(k^2 \pm 4 k\frac{\dot{\varphi}}{f} + m_{A'}^2\right)A'_{\pm} = 0
\end{equation}
For very light or massless dark photons, with no charged fermions, this effect dominates the friction mechanism and generates a background of non-thermal, very low frequency ($\sim \dot{\varphi}/f'$) vectors.  Using the work of \cite{Anber:2009ua}, the field velocity is roughly constant at $\dot{\varphi} = \xi H f'$, with $\xi\sim {\cal O}(100)$, corresponding to dark photon frequencies around a hundred times the Hubble rate.  The energy density in the dark photon approaches the vacuum energy $V$  when $C \sim V/(\xi f')$ \cite{Anber:2009ua}.  This friction will dominate over the thermal friction from the YM sector unless $f' \gg \sqrt{T/H} f$, a limit in which the dark photons don't thermalize anyway.  On the other hand, a full analysis of this version of dark energy radiation friction should run parallel to the inflation version of this model analyzed in \cite{Anber:2009ua}, and we suggest its cosmological analysis as dark energy for future work.

\subsection{Neutrinos}
\label{sec:neutrinos}

One very interesting possibility is the dark-energy thermalization of Standard Model neutrinos.  This could happen through the coupling of right-handed neutrinos to the dark energy radiation.  For example,  if fermions in the general Lagrangian \eqref{genericL} are in the adjoint representation of the non-Abelian group, this would allow for a simple way to thermalize neutral fermions, $N$, through the coupling
\begin{equation}
    {\cal L}\supset \frac{1}{f_N}G^a_{\mu\nu}\psi^a \sigma^{\mu\nu} N
\end{equation}
If $N$ is also a right-handed neutrino via the coupling $y h \ell N$, then it will thermalize the Standard Model neutrinos with masses $m_\nu \lesssim T$, potentially making them hotter than the standard temperature prediction for the CMB$\nu$ background.  While thermalization at late times will be generated as long as $f_N \ll $ TeV, this sector will be frozen out at early times and could easily have played a negligible role during
Big Bang Nucleosynthesis and recombination.

\section{Cosmological Signatures}
\label{sec:cosmology} 
Thermal friction has a unique cosmological signature through its modification of the late-time behavior of the dark-energy equation of state $w(a)$ as a function of the scale factor of the universe, $a$. The dynamics also gives rise to late-time spatial inhomogeneity (see \cite{Berghaus:2020new}) with small deviations from quintessence predictions, though these do not produce sizeable observables. However, the background level evolution has distinctive features from a cosmological constant $w_{CC}(a) = -1$, as well as other scalar field models, which may be observable with upcoming experiments. In this section we present the derivation of the thermal friction dark energy equation of state $w_{\text{DER}}(a)$.    

\begin{figure}[h]
    \centerline{\includegraphics[scale=0.55]{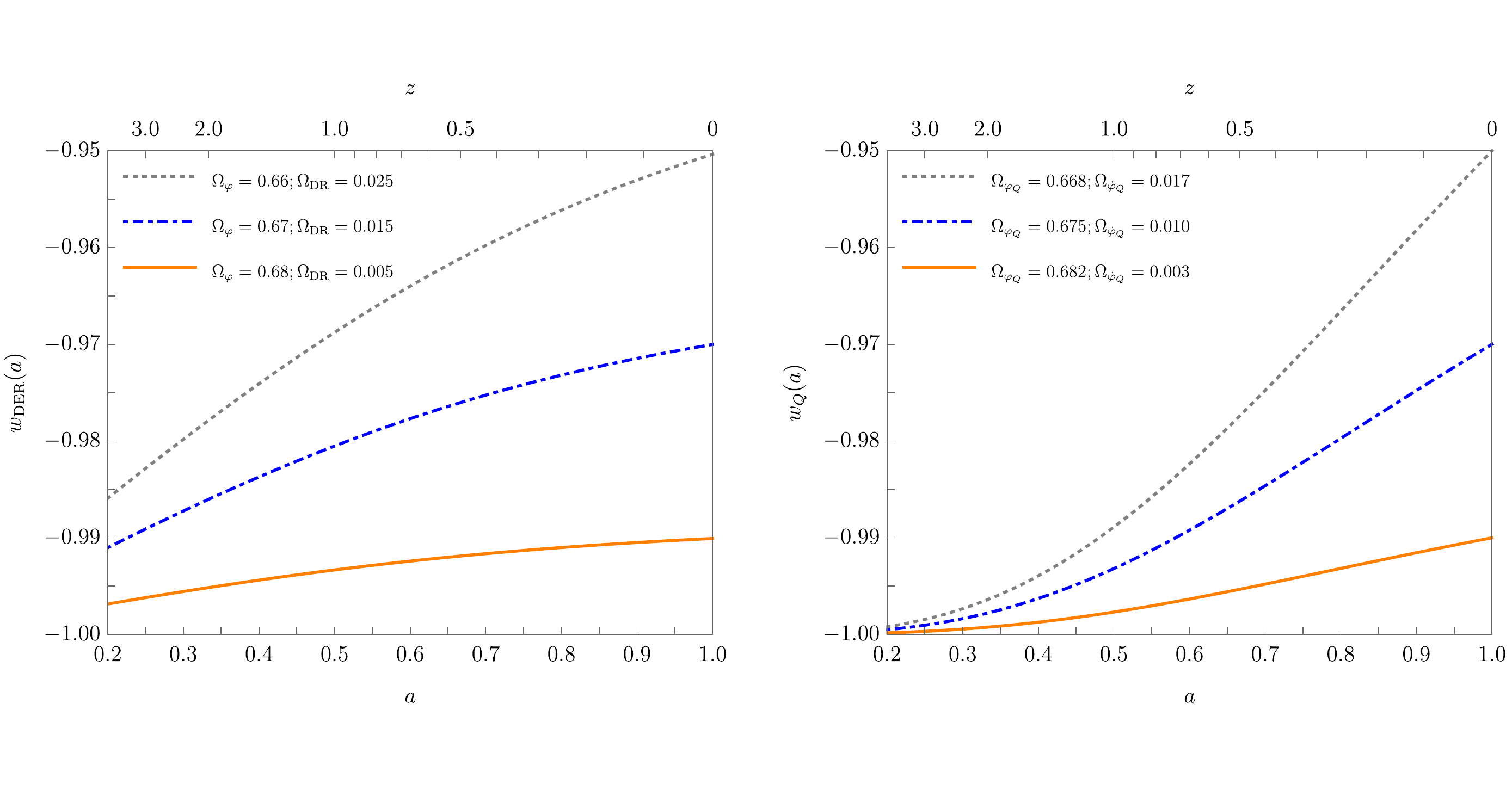}}
    \caption{The dark energy equation of state $w$ as a function of scale factor $a$ (redshift $z$) for thermal friction on the left and quintessence on the right. The abundances are defined as: $\Omega_{\varphi} = \frac{-C \varphi_0}{\rho_c}$, $\Omega_{\text{DR}} = \frac{\frac{\pi^2}{30}g_* T^4_0}{\rho_c}$, $\Omega_{\dot{\varphi}_Q} = \frac{\frac{1}{2}a_0^2 H^2_0 {\varphi'(a_0)}^2}{\rho_c}$ where $\rho_c$ is the critical density today. We assume $g_* = 7$ corresponding to a minimal thermal bath (comprised of three gauge bosons with two polarizations each and the axion with one scalar degree of freedom), and the best fit Planck value $\Omega_m = 0.315$ \cite{Aghanim:2018eyx}.}
    \label{EOS}
\end{figure}

The equations of motion of the scalar field-radiation system in equation \eqref{generalEOM} and \eqref{DRevolve} as a function of the scale factor $a$ are:
\begin{equation} \label{E1}
a^2 H^2 \varphi'' +\left(4a H^2 + a^2 H'H + a H \Upsilon \right)\varphi'  = C  
\end{equation}
\begin{equation} \label{E2}
\frac{4\pi^2}{30}g_{*}a H T'T^3 + \frac{4\pi^2}{30}g_{*}H T^4 = \Upsilon a^2 H^2{\varphi'}^2
\end{equation}
where primes denote derivatives with respect to $a$ and $g_*$ is the effective number of degrees of freedom in the dark radiation sector.  We will take $a=a_0\equiv 1$ to denote the scale factor today. Assuming the friction coefficient of a pure non-Abelian gauge group $\Upsilon = T^3/\tilde{f}^2$ (where any dependence on $\alpha$ and $N_c$ has been absorbed in $\tilde{f}$), one can now solve this system of equations for $\varphi(a)$ and $T(a)$ with appropriate boundary conditions.  However, focusing on the regime in which thermal friction is dominant ($H \ll T^3/\tilde{f}^2$) at all relevant times simplifies equations \eqref{E1} and \eqref{E2} to:
\begin{equation}\label{SEOM}
\varphi'(a) \approx \frac{\tilde{f}^2 C}{a H(a) T^3(a)}    
\end{equation}
\begin{equation} \label{REOM}
 T'(a)  + \frac{T(a)}{a} =   \frac{15 a H(a) {\varphi'(a)}^2}{2 \pi^2 g_* \tilde{f}^2}   
\end{equation}
where $g_*$ denotes the relativistic degrees of freedom in the radiation energy density $\rho_{\text{DR}} \equiv \frac{\pi^2}{30} g_* T^4$. We solve equation \eqref{REOM} perturbatively by approximating the Hubble function by its counterpart with a constant dark energy, $H(a) \approx  H^{(0)}(a) \equiv H_0 \sqrt{\frac{\Omega_m}{a^3}+ 1 -\Omega_m}$, where $\Omega_m$ is the ratio of matter density to critical energy density today (we are assuming a flat geometry).  In doing so, we obtain the following analytic expression for the temperature:
\begin{equation} \label{Toa}
T(a) = \left(\frac{\frac{15 C^2 \tilde{f}^2}{2 \pi^2 g_*} \sqrt{\frac{\Omega_m}{a^3}+ 1 -\Omega_m}  \left(8 a^3 (1- \Omega_m)-11 \Omega_m \left(1 -  {}_{2} F_1\left[1, \frac{4}{3};\frac{11}{6};-\frac{a^3(1-\Omega_m)}{\Omega_m} \right] \right) \right) }{8a^3 H_0 (1- \Omega_m)^2} \right)^{\frac{1}{7}} 
\end{equation}
which we plug into equation \eqref{SEOM} and numerically integrate to solve for $\varphi(a)$. Here, ${}_{2} F_1[a,b;c;z]$ is the Gaussian hypergeometric function.  In this regime the kinetic energy contribution is highly suppressed. Thus, in this limit $T(a)$ and $\varphi(a)$, and their values today $T_0\equiv T(a=1)$ and $\varphi_0\equiv \varphi(a=1)$, quantify  the dark-energy equation of state $w_{\text{DER}}$ as a function of $a$: 
\begin{equation} \label{EOSth}
w_{\text{DER}}(a) \simeq - 1 
\frac{- C \varphi(a) }{-C \varphi(a) + \frac{\pi^2}{30}g_* T^4(a)} + \frac{1}{3} \frac{\frac{\pi^2}{30}g_* T^4(a)}{-C \varphi(a) + \frac{\pi^2}{30}g_* T^4(a)}.
\end{equation}
 The value of the Hubble constant today, $H^2_0$, and the temperature today, $T_0$ (which could potentially be measured by direct detection discussed in Section \ref{sec:discussion}), determines the parameter combinations $C \varphi_0$ and $C \tilde{f}$, the only combinations appearing in the dark-energy equation of state itself (up to the discrete $g_*$). Since the constraints on $w_{\text{DER}}$ are inferred assuming a $\Lambda$CDM cosmology, a dedicated analysis is required to evaluate its bounds assuming a dynamic dark energy with thermal friction, which is an interesting prospect beyond the scope of the discussion presented here. For illustrative purposes, we plot $w_{\text{DER}}(a_0) \leq -0.95$ in Figure \ref{EOS} which corresponds to temperatures of about $T \lesssim 0.8 \, \text{meV}$, assuming $g_{*}=7$ (the case of $SU(2)$ with $\varphi$ also thermalized). Though there are three model parameters, the slope $C$, the value of the scalar field today $\phi_0$, and $\tilde{f}$ the friction strength parameter, there is no phenomenological sensitivity to all three parameters in this regime. This is due to $w_{\text{DER}}'(a_0)$ being dominated by $T'(a_0) \propto \left( C \tilde{f} \right)^\frac{2}{7}$ rather than $\varphi'(a_0)$ which does not lift the degeneracy between different combinations of $C$ and $\tilde{f}$. Thus the shape of the curves $w_{\text{DER}}(a,T_0)$ is fully determined by $w(a_0)$ as shown on the left of Figure \ref{EOS}.

In the opposite limit $H \gg \Upsilon$ ($\tilde{f}\rightarrow\infty$), we recover the familiar quintessence solution in which kinetic energy dominates and dark radiation is highly suppressed.  To solve for the quintessence field $\varphi_Q$ we turn off the thermal friction and recover the equation of motion of the regular quintessence scalar field whose terms were previously sub-dominant:  
\begin{equation} \label{EOMphiofa}
a^2 H^2 \varphi_Q'' +\left(4a H^2 + a^2 H'H \right)\varphi_Q'  = C   
\end{equation}
and find an equation of state:
\begin{equation}
w_Q(a) \simeq -1\frac{-C \varphi_Q(a)}{-C \varphi_Q(a) + \frac{1}{2}a^2 H(a)^2 {\varphi_Q'(a)}^2} + \frac{\frac{1}{2}a^2 H(a)^2 {\varphi_Q'(a)}^2}{-C \varphi_Q(a) + \frac{1}{2}a^2 H(a)^2 {\varphi_Q'(a)}^2}
\end{equation}
Using the Ansatz $H^{(0)}(a)$ (see text above equation \eqref{Toa}) we obtain the analytical solution for the quintessence field $\varphi_Q(a)$: 
\begin{equation} \label{phiQ}
\varphi_Q(a)  = \varphi_{Q,0} +\frac{2C\left(\text{tanh}^{-1}\left(\frac{\sqrt{1-\Omega_m}}{\sqrt{\frac{\Omega_m}{a^3}+1-\Omega_m}}\right)-\text{arctanh}^{-1}\left(\sqrt{1-\Omega_m} \right)\right)}{9H^2_0(1-\Omega_m)^{\frac{3}{2}}} 
\end{equation}
where $\varphi_{Q,0} \equiv \varphi_Q(a_0)$. To the right side of Figure \ref{EOS} we show the characteristic shape of quintessence curves with varying kinetic energy contributions. At early times $a < 0.3$, quintessence closely resembles a cosmological constant. Only recently did the field obtain a sizeable kinetic energy component $\Omega_{\dot{\varphi}}$ whose total amount is smaller than the amount of dark radiation in the thermal friction model that results in the same value for $w(a_0)$ today. On the contrary, the change of $w_{\text{DER}}(a)$ in the thermal friction model is more gradual. Even at early times there exists a sizeable deviation from a cosmological constant. The slowly rising temperature increases this deviation which asympotatically approaches a constant value. Quintessence shows a differing trend with the deviation from a cosmological constant linearly growing. Quintessence also requires much larger field values venturing into the trans-planckian field regime in order to remain within $w(a_0) < -0.95$, whereas thermal friction has a wide set of viable parameters that are sub-Planckian. For example the gray dashed line on the left of Figure \ref{EOS} corresponds to $\varphi(a_0) \sim - 0.01 \left(\frac{\tilde{f}}{1 \text{ GeV}} \right) M_{pl}$, whereas the gray dashed line on the right of Figure \ref{EOS} corresponds to $\varphi_Q(a_0) \sim -2 M_{pl}$. This is due to the thermal friction being much larger than Hubble friction which allows for much steeper slopes within the experimental constraints leading to smaller field values. The perturbative descriptions of the two regimes discussed in this section are in very good agreement with exact numerical solutions of \eqref{E1} and \eqref{E2},  with deviations smaller than $0.02\%$ for all curves shown in Figure \ref{EOS}.

As we have seen above, in our scenario there is actually only a single additional parameter beyond the usual $\Lambda$CDM parameters which will determine our entire equation of state as a function of time, $w_{DER}(a)$, at leading order.  Further, this also determines the temperature of the dark radiation today, which could be separately measured in a direct detection experiment.
If we find evidence that dark energy is not a cosmological constant then we will immediately want to measure its variation as a function of cosmic time.
If the (even roughly) measured $w(a)$ curve fits with our constrained family of predictions, it would be a striking success of the model.  Further, it would then make a sharp prediction for the temperature of the dark radiation which would start a race to directly detect this radiation in the lab.

\section{Discussion and Conclusion}
\label{sec:discussion}
The mechanisms discussed above can produce a thermalized local energy density in dark radiation that can be as large as a few percent of the dark energy density $\sim \text{meV}^4$ today, much larger than the energy density in the CMB.  This radiation would have a potentially observable effect on the expansion rate of the universe, distinguishable from just a rolling scalar field (quintessence).  Interestingly, even though most of the kinetic energy of the dark energy field has gone into radiation, the usual cosmic birefringence signal could still be observed at the same level.  The radiation might also be directly detectable.  This radiation may be in the form of pure Yang Mills gluons, hidden photons, right handed neutrinos or milli-charged particles. It would be interesting to develop experimental methods to directly detect these particles, similar to dark matter direct detection efforts.

Of these types of radiation, a pure Yang Mills sectors is the most difficult to detect since it can only interact with the nucleons and electrons through operators of high dimension. 
The other portals are more promising. There are many experimental efforts \cite{Chaudhuri:2014dla,Bunting:2017net, Carosi:2020akt, Knapen:2017ekk,Essig:2019kfe,Arvanitaki:2017nhi, Chen:2020jia} currently underway to detect hidden photons. These experiments are focused on detecting cold dark matter  at lower frequencies, with the experiment utilizing the expected narrow-band nature of the dark matter signal to resonantly boost sensitivity. But, there are emerging attempts to detect heavier hidden photons in the meV range. In this energy range, due to the difficulty of creating electromagnetic resonators, dark matter direct detection efforts are also broadband devices. Moreover, the fundamental sensing technology necessary to make progress in this energy range is the technology necessary to detect single photons in the THz band. Due to the practical importance of this technology, it is likely that there will be continued investment in developing these sensors. This will lead to direct improvement in our ability to search for hidden photon dark radiation produced by dark energy. Similarly, experiments have also been proposed to detect milli-charged dark matter. It would be interesting to evaluate the reach of these experiments to relativistic milli-charged particles. While relavitistic particles will spend less time transiting a detector than non-relativistic dark matter, they also interact more strongly with magnetic fields and this could potentially be used to create magnetic focusing devices to boost the experimental reach. We leave a detailed discussion of this topic for future work. 

The case of right handed neutrinos is particularly interesting. At the energy scale $\sim$ meV relevant to dark energy radiation, the right handed neutrino easily mixes with the left handed neutrino and can thus interact with the standard model through the weak interactions. The right handed neutrinos produced from dark energy could thus have a much larger energy density (by $\sim$ four orders of magnitude) than the standard model (left handed) neutrino background. Due to  strong mixing, these right handed neutrinos can also be searched for using experiments such as PTOLEMY
\cite{Baracchini:2018wwj} that are dedicated to probing the cosmic neutrino background. In these experiments, the dark energy produced neutrinos may have an energy density 1000x larger than the standard model neutrino background giving rise to an enhanced rate in the experiment. Moreover, these neutrinos could also be at higher energies $\sim$ meV as opposed to $\sim 0.1$ meV expected from the cosmic neutrino background. The increased energy should help in detection in an experiment like PTOLEMY that searches for a kinematic end point. 

A discovery of meV scale radiation in an experiment would be proof that this radiation was produced at late times in the universe well after the CMB. The two contenders for the source of such a signal would be dark matter and dark energy.  While the meV scale would be indicative of dark energy, it will be necessary to develop experimental methods to distinguish these two possibilities.
The first check would be to probe the equation of state of dark energy through cosmological measurements. With the advent of gravitational wave detectors that offers the tantalizing possibility of genuine standard candles, it might be possible to extend the reach of such cosmological probes.   Further, once it is known that the radiation couples to the Standard Model, then the scalar dark energy field giving rise to the radiation must couple as well at some level.  Interestingly, this rolling scalar field could then give a detectable cosmic birefringence signal in the CMB, which would be another indication for this scenario.

This paper has focused on the conversion of dark energy into thermal radiation through technically natural couplings between a rolling field and a gauge sector. These kinds of couplings can also source another class of signals.  Tachyonic instability \cite{Anber:2009ua} can result in the conversion of the kinetic energy of the field into ultra-long ($\sim 100/H_0$) wavelength modes of a vector field. Canonical examples of such vector fields are $B-L$ gauge bosons and a hidden photon. Signals from the former can potentially be looked for using torsion balances/accelerometers\cite{Graham:2015ifn}. The latter case may be phenomenologically interesting - given the plasma mass of the photon, the hidden photon field is likely to manifest itself as a spatially coherent magnetic field in the universe and it might be interesting to see if such a magnetic field could be the source of the observed long range coherence in extra-galactic magnetic fields \cite{Kamada:2018kyi}.

The seriousness of the cosmological constant problem and the most reasonable approach to its solution makes a strong case for dynamical dark energy.
Dark matter also presents a deep mystery about the contents of our universe.
While dark matter was discovered through its gravitational effects, theoretical motivations suggested a robust experimental program to directly detect it in the laboratory.
Dark energy too was discovered through its gravitational impact on our universe.
In this paper we have argued that theoretical considerations also motivate a broader experimental program to directly detect a dynamical component of dark energy in the laboratory.

\section*{Acknowledgments}

 DK and SR are supported in part by the NSF under grant PHY-1818899.  SR acknowledges travel support from the Gordon and Betty Moore Foundation and the American Physical Society enabling him to visit Stanford to complete this work.  SR is also supported by the DoE under a QuantISED grant for MAGIS. KB was supported by in part by the NSF under grant PHY-1818899 and PHYS-1915093, and the NASA award NNX17AK38G.  SR and PWG are also supported by the SQMS quantum center. PWG  acknowledges the support provided by NSF Grant PHY-1720397, the Heising-Simons Foundation Grant 2018-0765, the DOE HEP QuantISED award 100495, and the Gordon and Betty Moore Foundation Grant GBMF7946.

\bibliographystyle{unsrt}
\bibliography{refs}

\end{document}